\documentclass[11pt]{article}
\usepackage[english]{babel}
\input{epsf}
\usepackage{mathtext}
\usepackage{graphicx}
\usepackage{graphics,dcolumn,bm,float}
\usepackage{amssymb,amsmath,rotate,color}
\usepackage{times}
\usepackage{tabularx}

\newcommand{\hhh}{{\cal H}}

\newcommand{\TT}{{\cal T}}
\newcommand{\CC}{{\cal C}}
\newcommand{\CA}{{\cal A}}
\newcommand{\CB}{{\cal B}}
\newcommand{\CD}{{\cal D}}
\newcommand{\CE}{{\cal E}}
\newcommand{\CF}{{\cal F}}
\newcommand{\CP}{{\cal P}}
\newcommand{\be}{\begin{equation}}
\newcommand{\ene}{\end{equation}}
\newcommand{\ba}{\begin{array}}
\newcommand{\ea}{\end{array}}

\begin{document}

\title{Tuneable superconducting effective gap in graphene-TMDC heterostructures}
\author{S.F. Ebadzadeh$^{a}$, H. Goudarzi$^a$\footnote{Corresponding author; e-mail address: h.goudarzi@urmia.ac.ir}, M. Khezerlou$^{a,b}$\\
\footnotesize\textit{$^a$Department of Physics, Faculty of Science, Urmia University, P.O.Box: 165, Urmia, Iran}\\
\footnotesize\textit{$^b$National Elites foundation, Iran}}
\date{}
\maketitle

\begin{abstract}
Growth of graphene on monolayer transition-metal dichalcogenides presents opening on band gap and giant spin-orbit coupling which paves the way to achieve a useful hybrid structure for electronics and spintronics applications. Increase of the atomic number of transition-metal results in a large SOC, where eventually a band inversion appears in graphene-$WSe_2$. We consider superconductor induction by proximity effect to the graphene-TMDC hybrid structure. As a necessity of formalism, we introduce a proper time-reversal and particle-hole symmetry operators, under which the $8 \times 8$ Dirac-Bogoliubov-de Gennes low-energy effective Hamiltonian is invariant. Resulting superconducting electron-hole excitations shows that, the essential dynamical parameters $\lambda_I^{A,B}$ and $\lambda_R$ have significant effect on superconducting excitations and, specifically, subgap energy. Dependence of the superconducting energy excitation on chemical potential is explored. The signature of spin triplet $p$-wave pairing symmetry in the system is found to increase the subgap superconducting energy, in comparing to $s$-wave symmetry.
\end{abstract}

\section{Introduction}
The enormous theoretical and experimental probes have been done on graphene since, it was produced as two dimensional material in 2004 \cite{KS,A}. Although, graphene possesses exotic properties such as high carrier mobility, low resistivity, more optical absorption at the plasmon-resonance (see Ref. \cite{S}), and etc., the gapless quasiparticle electronic structure makes it to present no potential electronics applications. Instead of it, the liner dispersion in Dirac point gives rise to detect interesting physical phenomena, such as Klein tunneling \cite{MI} and spin-valley resolved conductance \cite{3,2}. As an another important feature of 2D materials, the spin-orbit coupling (SOC) in graphene is achieved to be very small$\approx 1\mu eV$ \cite{C}, which can be considered as a weakness. Taking into $d$-orbital contribution to SOC can improve it up to $\approx 24\mu eV$ \cite{M}. Actually, there are several ways to enhance the size of SOC in graphene, which is potentially suitable for spintronic applications \cite{D}, for example, by adding hydrogen or fluorene atoms (leading to increasing $sp^3$ hybridization \cite{AJ}), by decorating heavy adatoms \cite{CW} and using different substrates \cite{G}. One of the effective possibility to induce a sizeable SOC is the use of a layered transition-metal dichalcogenide (TMDC) as substrate \cite{K,J,MG,Z,F}.

However, a monolayer TMDC (which transition metal can be $Mo$ or $W$ atom and $S$ or $Se$ can denote chalcogen atom) presents a significant spin-orbit interaction in its band structure. Moreover, its honeycomb-like lattice structure provides possibility to dynamically hybridize with graphene. Indeed, the experimental growth of graphene on $MoS_2$ or $WS_2$ layer was very successful, see Refs. \cite{Z,CP,AA,H,JR}. There are several essential applications of such hybrid structures used in a basis for nonvolatile memory \cite{B}, sensitive photodetection \cite{CC}, and gate-tunable persistent photoconductivity \cite{R}. Graphene grown on ML-MDS presents a large SOC $\approx 1 meV$ \cite{F}. Importantly, the spin Hall effect in graphene has been appeared at room temperature resulted from proximity-induced SOC via a few-layer $WS_2$ deposition \cite{AA}. In this structure, the linear-in-momentum band structure is preserved within the TMDC direct band gap, and Dirac point shifts toward the valence band with increasing the atomic number of the chalcogen. The increasing atomic number of the transition metal results in a stronger SOC, while Dirac band structure remains conventional. Whereas, graphene-$WSe_2$ hybrid shows a band inversion \cite{MG}. In these systems, the appearance of spin-resolved different gaps around Dirac point is originated from the fact that the spatial symmetry in graphene lattice (i.e. sublattice symmetry) is reduced. Also, existence of sizeable SOC can be responsible to the tunable Dirac band gap. The size and topological nature of the gaps are determined by the interplay between sublattice symmetry breaking and enhanced SOC. In Ref. \cite{SE}, the authors show that graphene-$MoS_2$ hybrid can demonstrate a quantum spin-Hall insulator or valley-Hall insulator, depending on SOC dominates the sublattice symmetry breaking or vice versa. However, SOC removes the degeneracy of $K$ and $K'$ valleys, and sublattice symmetry breaking leads to spin-resolved transport.

Indeed, breaking the pseudospin symmetry in graphene results in opening an orbital band gap. This can be realized by the fact that the atoms in sublattices A and B distinctly interact with effective potential of ML-TMDC \cite{MG,F}. The orbital band gap is retained even in the absence of SOC \cite{F}. The spin degeneracy of the valence and conduction bands is removed by SOC. As a key feature of these  dynamically complicated hybrid structures, the superconducting order induction by proximity effect can lead to arising electron-hole low-energy excitations, which can be fundamentally distinct from the similar situation in alone two-dimensional structures. Specifically, particle-hole and time-reversal symmetries requirement needs to reconsider in order to Dirac quesiparticles satisfy the Bogoliubov-de Gennes (BdG) formalisms. In recent few years, the superconducting electron-hole excitations in graphene and monolayer molybdenum disulfide (ML-MDS) has been separately investigated in several works \cite{3,2,JB,MH,BZ}. 
 
If the $s$-wave superconductivity is induced in ML-MDS and graphene, the cooper pairs possess opposite spins and form excitations from  different valleys \cite{BZ}. Consequently, the net spin and momentum is not carried by the cooper pairs, since the time reversal symmetry is preserved. The superconductivity spectrum is similar to the Mexican hat in graphene and ML-MDS. 
In this paper, we consider spin-singlet and spin-triplet pair potential components in the effective low-energy Hamiltonian of graphene-TMDC heterostructures, (see, a schematic sketch of structure in Fig.1). We intend to investigate the time-reversal and particle-hole symmetry operators for corresponding Dirac and BdG Hamiltonian, respectively. We study, in Sec. 2, the effective Hamiltonian with low-energy electronic structure, which is achieved by using the tight-binding approach for superlattice of graphene-TMDC. To present the hole dispersion energy, the second block of Hamiltonian is obtained by the time-reversal symmetry invariant. Section 3 is devoted to unveil the electron-hole superconducting excitations when a superconductor $s$-wave or $p$-wave order is induced by proximity effect in the heterostructure. In the last section, the main characteristics of proposed structure are summarized.

\section{Effective Hamiltonian}

The Brillouin zone of graphene-ML-MDS is hexagonal and the low energy fermionic excitations treat as massive Dirac particles around the edges of this zone.
Using the tight-binding formalism provides an effective method to investigate dynamical properties of graphene-ML-MDS heterostructure. The tight-binding Hamiltonian of  graphene-ML-MDS contains all symmetry nearest- and next-nearest-neighbor to fully maintain the sublattice inversion asymmetry. The broken inversion symmetry in this system originates from the proximity of the ML-MDS with graphene, which the Rashba SOC term is appeared in the effective Hamiltonian \cite{MM}. Considering the hybridization of the graphene orbitals with $d$ orbitals $(d_{z^2 }, d_{xy}$ and $d_{x^2-y^2 })$  of ML-MDS, the sublattice-resolved and giant spin-orbit couplings appear in Hamiltonian. This Hamiltonian consists of orbital and spin-orbit parts $H=H_{orb.}+H_{so.}$. The orbital Hamiltonian is described as
\begin{equation}
H_{orb.}=\sum_{\left\langle i,j\right\rangle,\sigma}tc^{\dagger}_{i \sigma}c_{j \sigma}+\sum_{i,\sigma}\Delta\xi_{c_i)}c^{\dagger}_{i \sigma}c_{i \sigma},
\end{equation}
which refers to govern dynamics on the orbital energy. The hopping $t$  denotes the nearest-neighbor interaction, and $\left\langle ij\right\rangle$ denotes the sum over nearest-neighbor sites. The parameters $c^{\dagger}_{i \sigma}\equiv(a^{\dagger}_{i \sigma},b^{\dagger}_{i \sigma})$ and $c_{i \sigma}\equiv(a_{i \sigma},b_{i \sigma})$ are the creation and annihilation operators, respectively, for an electron with spin $\sigma\left(\uparrow,\downarrow\right)$ at site $i$ belonging to the sublattice A or B. The quantity $\Delta$ refers to the energy difference perceived by atoms in the sublattices A $\left(\xi_{a_i}=1\right)$ and B $\left(\xi_{b_i}=-1\right)$. The spin-orbit Hamiltonian is given by
\begin{eqnarray}
H_{so.}=\frac{2i}{3}\sum_{\left\langle i,j\right\rangle}\sum_{\sigma,\acute{\sigma}}c^{\dagger}_{i \sigma}c_{j\acute{\sigma}}\left[\lambda_{R}\left(\hat{\mathbf{s}}\times \hat{d}_{ij}\right)_{z}\right]_{\sigma\acute{\sigma}}+\frac{i}{3\sqrt{3}}\sum_{\ll i,j\gg}\sum_{\sigma,\acute{\sigma}}c^{\dagger}_{i \sigma}c_{j\acute{\sigma}}\lambda_{I}^{ci}v_{ij}\left(\hat{s}_{z}\right)_{\sigma\acute{\sigma}}\nonumber\\
 +\frac{2i}{3}\sum_{\ll i,j\gg}\sum_{\sigma,\acute{\sigma}}c^{\dagger}_{i \sigma}c_{j\acute{\sigma}}\left[\lambda_{PIA}^{ci}\left(\hat{\mathbf{s}}\times\hat{D}_{ij}\right)_z\right]_{\sigma\acute{\sigma}}.
\end{eqnarray}
The first term describes Rashba spin-orbit coupling with strength $\lambda_R$. $\hat{s}$ denotes a vector of Pauli matrices that acts on spin space. $\hat{d}_{ij}$ is the unit vector pointing from site $j$ to site $i$ that connects the nearest neighbor atoms A and B. The second term is the intrinsic SOC parameterized by $\lambda_I^{(c_i)}=\lambda_I^A(\lambda_I^B)$ for sublattice A(B). Here, $v_{ij}=1(-1)$ indicates the clockwise(anticlockwise) hopping path and $\ll i,j\gg$ is the summation over the next-nearest-neighbor hopping. The final term describes pesudospin inversion asymmetry (PIA) concerning to SOC where $\lambda_{PIA}^{(c_i)}=\lambda_{PIA}^A(\lambda_{PIA}^B)$ refers to coupling intensity for sublattice A(B). The unit vector pointing from site $j$ to its next-nearest-neighbor at site $i$ is denoted by $\hat{D}_{ij}$.

However, the intrinsic and PIA SOC is resulted from the breaking sublattice (pseudospin) symmetry. Eventually, low-energy Hamiltonian around the $K$ and $K'$ Dirac points is given by its  Fourier transformation as follows \cite{MG}
\begin{equation}
H_{eff.}^{K(K')}=H_1+H_2+H_3 ;
\end{equation}
$$
H_1= v_{F}\hbar\left(\tau\hat{\sigma}_{x}k_{x}+\hat{\sigma}_{y}k_{y}\right)+\Delta\hat{\sigma}_{z},
$$
$$ H_2=\frac{1}{2}\left(\lambda_{I}^{A}\left(\hat{\sigma}_{z}+\hat{\sigma}_{0}\right)+\lambda_{I}^{B}\left(\hat{\sigma}_{z}-\hat{\sigma}_{0}\right)\right)+\lambda_{R}\left(\tau\hat{\sigma}_{x}\hat{s}_{y}-\hat{\sigma}_{y}\hat{s}_{x}\right),
$$
and
$$
H_3=\frac{a}{2}\left(\lambda_{PIA}^A\left(\hat{\sigma}_{z}+\hat{\sigma}_{0}\right)+\lambda_{PIA}^B\left(\hat{\sigma}_{z}-\hat{\sigma}_{0}\right)\right)\left(k_{x}\hat{s}_{y}-k_{y}\hat{s}_{x}\right).
$$
The term $H_1$ describes gapped Dirac state and $H_2$ contains intrinsic and Rashba types spin-orbit coupling parts. The $H_3$ term consists of pesudospin inversion asymmetry spin-orbit coupling. $\hat{\sigma}_i (i=0,x,y,z)$ denotes the Pauli matrices which operate on the pseudospin sublattice A and B space. The valley index $\tau=+1(-1)$ denotes valley $K(K')$, and $v_F=\frac{\sqrt{3}at}{2\hbar}$ is Fermi velocity of the quasiparticles. The bond parameter $a=2.46 {\buildrel_{\circ}\over{\mathrm{A}}}$ refers to the lattice constant of pristine graphene. Hence, resulting effective Hamiltonian of graphene-ML-MDS is given as follows:
$$
H_{eff.}^{K(K')}(k_x,k_y)=
$$
\begin{equation}
\left(\begin{array}{cccc}
\Delta+\tau\lambda_{I}^{A}&a\lambda_{PIA}^{A}(-ik_{x}-k_{y})&v_{F}\hbar(\tau k_{x}-ik_{y})&\lambda_{R}(-i\tau+i)\\
a\lambda_{PIA}^{A}(ik_{x}-k_{y})&\Delta-\tau\lambda_{I}^{A}&\lambda_{R}(i\tau+i)&v_{F}\hbar(\tau k_{x}-ik_{y})\\
v_{F}\hbar(\tau k_{x}+ik_{y})&\lambda_{R}(-i\tau-i)&-\Delta-\tau\lambda_{I}^{B}&a\lambda_{PIA}^{B}(ik_{x}+k_{y})\\
\lambda_{R}(i\tau-i)&v_{F}\hbar(\tau k_{x}+ik_{y})&a\lambda_{PIA}^{B}(-ik_{x}+k_{y})&-\Delta+\tau\lambda_{I}^{B}
\end{array}\right).
\end{equation}
The graphene-ML-MDS Hamiltonian is the same as the other Hamiltonian of graphene-TMDC. The electronic structure parameters of graphene-TMDC heterostructures for two composition types of transition metals Mo and W with chalcogens S and Se are listed in table 1.

It is noticed that, on the honeycomb lattice, like graphene-TMDC, the proper time-reversal symmetry operator may result in interchange between two valleys. It really leads to an important result, e.g. Andreev reflection is related to coupling two different valleys in graphene \cite{JB}. Therefore, introducing a form of time-reversal symmetry to meet the requirement $\hat{\tilde{\TT}}H_{eff.}^{K*}\hat{\tilde{\TT}}^{-1}=H_{eff.}^{K'}$, for which $4\times 4$ Hamiltonian belonging to $K$-valley transforms into that in $K'$-valley, yields:
$$
\hat{\tilde{\TT}}=-(\hat{\sigma}_z\otimes\hat{\sigma}_x)\hat{\CC},
$$
where $\hat{\CC}$ denotes the complex conjugate operator. Diagonalizing Eq. (4) leads to the electronic band structure 
\begin{equation} 
\CE=\frac{\iota}{\sqrt{2}}\sqrt{\left(\Delta^2+\CA+\CB\right)+\eta\sqrt{\left(\Delta^2+\CA+\CB\right)^2-4\CD}},
\end{equation}
where 
$$\CA=\Delta^2+4\lambda_{R}^{2}+2(\left(v_{F}\hbar\left|k\right|\right)^2,$$ $$\CB=\left(\lambda_{I}^{A}\right)^2+\left(\lambda_{I}^{B}\right)^2,$$ $$\CD=4\lambda_{R}^2\CF+\Delta^2\left(\CA-\CB\right)+\left(\left(v_{F}\hbar\left|k\right|\right)^2+\lambda_{I}^{A}\lambda_{I}^{B}\right)^2,$$
and
$$\CF=\Delta\left(\lambda_{I}^{A}-\lambda_{I}^{B}\right)-\lambda_{I}^{A}\lambda_{I}^{B}.$$
The parameter $\iota=+1(-1)$ illustrates the conduction(valance) band, and $\eta=\pm 1$ specifies the up and down spin subbands of the conduction and valence bands. It is Notice that we neglect PIA spin-orbit coupling, because its effect on the low-energy band structure is very small.

 With above symmetry consideration, we are now able to construct an effective Hamiltonian describing the relativistic electron-like and hole-like quasiparticles. The $8\times 8$ effective Hamiltonian describing electron-like and hole-like Dirac fermions yields:
\begin{equation}
\hhh(k)=\left(\begin{array}{cc}
H_{eff.}^{K}&0\\
0&\hat{\tilde{\TT}}H_{eff.}^{K*}\hat{\tilde{\TT}}^{-1}
\end{array}\right).
\end{equation}
This Hamiltonian needs to conserve the time-reversal symmetry, which is given by $\hat{\TT}\hhh(k)\hat{\TT}^{-1}=\hhh^{*}(-k)$. We find a proper time-reversal  symmetry invariant of Hamiltonian (6) to be provided by introducing the operator
$$
\hat{\TT}=\left(\hat{\sigma}_{y}\otimes\hat{\sigma}_{0}\otimes\hat{\sigma}_{y}\right)\hat{\CC}.
$$
In Figs. 2 the dispersion energy of graphene-TMDC belonging to two sublattices A and B is presented, where resulting sizeable spin-splitting is appeared in both conduction and valence bands in graphene band structure. Since SOC is encouraged from transition metal Mo (Figs. 2(a) and 2(b)) to W (Figs. 2(c) and 2(d)) atom, corresponding band spin-splitting also is enhanced. Dirac point energy gap strongly depends on the type of transition metal. Specifically, in graphene-tungsten disulfide hybrid structure the band gap is almost closed for opposite subband spin, that it exhibits a semimetal phase. In Fig. 2(d), we show an inversion band regime around Dirac point for graphene-tungsten diselenide hybrid, owing to large intrinsic SOC, comparing to the normal band gap of TMDC, where $\left|\lambda_I^A+\lambda_I^B\right|>\Delta$.
 
\section{Superconducting graphene-TMDC}

As a main point of this paper, we proceed to consider proximity-induced superconductor pairing order in the graphene-TMDC monolayer. Noticed that, in the last decade, superconducting order induction to the Dirac-like fermions in garphene or ML-MDS has been reported by many authors, experimentally \cite{4,5,6,7,8,DC} and theoretically \cite{3,2,JB,MH,BZ,9}. Heersche, et al. put graphene layers in contacted with superconducting electrodes consisting of a titanium-aluminium bilayer, and reported a systematic research of superconducting induction in graphene layers \cite{4}. Recently, the gate-induced superconductivity in monolayer $MoS_2$ has been investigated by performing tunneling spectroscopy to specify the energy-dependence density of state \cite{DC}. Considering superconducting gap induction by proximity effect, the low energy quasiparticle excitations can be described by the Dirac-Bogoliubov-de Gennes (DBdG) Hamiltonian: 
\begin{equation}
\hhh_{DBdG}(k)=\left(\begin{array}{cc}
\hat{H}_{eff}^{K(K')}-\mu_S & \hat{\Delta}_S(\mathbf{k})\\
\hat{\Delta}_S ^{*}(\mathbf{k})& \mu_S-\hat{H}_{eff}^{K(K')}
\end{array}\right),
\end{equation}
where $\mu_S$ is the chemical potential. The superconducting pair potential $\hat{\Delta}_S(k)$ couples the electron excitation in one valley with hole excitation in the other valley. Here, the superconducting pair potential is considered to be spin-singlet $s$-wave or spin-triplet $p$-wave symmetry. It is of more importance that despite induction of conventional $s$-wave symmetry, the triplet superconducting components can be created in the materials with strong spin-orbit coupling.
 
\subsection{$s$-wave symmetry}

The order parameter of superconductor is define by a function of momentum $(\mathbf{k}, \mathbf{-k})$ and spin $(s,s')$.
In $s$-wave paring (spin singlet), the superconducting pair potential may be written $\hat{\Delta}_S(\mathbf{k})=\Delta_{0}e^{i\varphi}\hat{\sigma}_{0}$, where $\Delta_{0}$ denotes the value of the superconductor gap and $\varphi$ is the phase of the pair potential of superconducting. The $s$-wave superconductor gap matrix includes diagonal singlet components.
It needs to introduce a $8\times 8$ DBdG Hamiltonian, which may necessarily be invariant under the particle-hole symmetry
$$
\hat{\CP}\hhh_{DBdG}(k)\hat{\CP}^{-1}=-\hhh^{*}_{DBdG}(-k).
$$
To this end, we introduce following particle-hole symmetry operator:
$$
\hat{\CP}=\left(\hat{\sigma}_{y}\otimes \hat{\sigma}_{z}\otimes \hat{\sigma}_z\right)\hat{\CC}.
$$

By diagonalizing Hamiltonian (7), we obtain the superconducting electron-hole excitations in graphene-TMDC heterostructures, and, of course, resulting appeared superconducting subgap around Fermi surface. Having this situation, one can be able to consider specular Andreev reflection and resulting subgap conductance in a related junction. This can be fundamentally distinct from similar situation with alone graphene, which has neither Dirac gap nor intrinsic large SOC. We study the effect of intrinsic $\lambda_I^{A,B}$ and Rashba $\lambda_R$ types SOC on the superconducting excitations of two different group of transition metal dichalcogenides $(MoS_2, MoSe_2)$  and $(WS_2, WSe_2)$. Actually, above essential parameters play a crucial role in graphene-TMDC hybrid structures. When used transition metal is Mo atom, the absence of intrinsic and Rashba SOC term, in leads to increasing Dirac band gap and almost vanishing superconducting gap in electron-hole excitations, as shown in Figs. 3(a) and 3(b). Rashba type SOC has almost no influence in energy excitations, and in the case of $\lambda_I^{A,B}=0$, the band gap is increased. Hence, it is expected to suppress Andreev reflection, originated from excitations below superconducting subgap, in the lack of SOC.

When Mo atom is replaced by tungsten atom in superconducting monolayer TMDC, we find two distinct features in electron-hole excitations. First, the superconducting effective subgap is significantly increased, as seen in Figs. 3(c) and 3(d), in comparing to those appeared in system with Mo atom. Secondly, Fig. 3(c) shows that, the absence of intrinsic and also Rashba spin-orbit couplings gives rise to strongly increasing Dirac gap and, of course, vanishing the superconducting gap. As a key result, this can be originated from the fact that the triplet component superconducting correlations is a dominant procedure in the presence of strong SOC.

Finally, graphene-$WSe_2$ hybrid shows interesting band structure in proximity with a superconductor order. The inversion band regime, which was observed in its low-energy dispersion (Fig. 2(d)), is preserved. This feature makes the possibility to dynamically occur electron-hole pair coupling twice for the two different Fermi wavevectors. In contrast to the other three hybrid structures, which there is hole excitations near zero wavevector, we find to appear electron-like excitations around $k=0$, (see, Fig. 3(d)). Consequently, as shown in Fig. 4, the Mexican hat shape of superconducting conduction band excitations in $MoS_2$, $MoSe_2$ and $WS_2$ (see, Ref. \cite{10}) is converted to Cowboy hat in superconductor graphene hybrid with $WSe_2$.
Particularly, for $|k|\cong 0.2$, the conduction and valence bands touch each other, when Rashba SOC term is neglected. This means that gapless state is formed for a wavevector, which is different from Fermi wavevector $|k_F|\cong 3.5$, see the green dashed line in Fig. 3(d).
Also, in contrast to the system with $WS_2$, the Dirac point energy gap remains constant, when all SOC terms are zero $\lambda_I^A=\lambda_I^B=\lambda_R=0$, and, importantly, double conduction band electron-hole excitations is converted to the common single electron-hole excitation.

Furthermore, we proceed to investigate the signature of chemical potential of superconductor in energy excitations. Actually, there is a physical necessity of superconductor chemical potential to be much larger than pair potential, $\mu_S\gg \Delta_0$. In Figs. 5, we observe that Fermi wavevector, in which the superconducting effective gap occurs, is displaced. This can be an expected consequence, because the Fermi surface of superconducting excitations depends on the chemical potential.  

\subsection{$p$-wave symmetry}

The unconventional triplet superconductivity induction to the 2D structures, like graphene or ML-MDS, can be possible by using $Sr_2RuO_4$ material \cite{DG}. The effect of $p$-wave paring symmetry on the superconducting excitation has been theoretically investigated in graphene and $MoS_2$ Ref. \cite{MH,11}, where, in monolayer $MoS_2$, the electron-hole excitations has been found to be semi-gapless. This feature shows that the superconducting subgap weakens, leading to suppressing more or less the Andreev reflection process. Manifestly, the spin part of triplet order parameter is even spin channel, while the spatial part is an odd function (orbital number $l=1$). The superconducting pair potential for $p$-wave pairing is defined using the $\mathbf{d}(\mathbf{k})$ vector as
$$
\hat{\Delta}_S(\mathbf{k})=\left[\mathbf{d}(\mathbf{k})\cdot\hat{\mathbf{s}}\right]i\hat{s}_{y},
$$ 
where $\mathbf{d}\left(\mathbf{k}\right)$ is odd-parity function of $\mathbf{k}$. The gap matrix $(\hat{\Delta}_S(\mathbf{k}))$ of $p$-wave paring is off diagonal triplet components, which has an essential distinction from conventional singlet. The operator of particle-hole symmetry of $p$-wave paring is the same as in $s$-wave paring.

We find the behavior of superconducting excitations with $p$-wave symmetry to be the same as in $s$-wave symmetry. But, comparing the superconducting effective subgap in $s$ and $p$-wave cases shows that it increases when the proximity-induced pair potential is the unconventional spin-triplet $p$-wave. To unveil this effect, we plot the energy eigenvalue of Hamiltonian (7) as a function of electron-hole wavevector for a large magnitude of superconductor pair potential $\Delta_0=0.2 \ meV$, as shown in Figs. 6(a) and (b). Indeed, the interplay between large amount of SOC in TMDC monolayers and non-zero orbital quantum number $(l=1)$ of $p$-wave order causes the increase of effective superconductor gap. As pointed out previously, according to the tendency of appearance of triplet components with inducing an $s$-wave order in materials with large SOC, inducing an intrinsic spin-triplet $p$-wave superconductor, in fact, results in a higher superconducting effective subgap in graphene-TMDC hybrid structure, in comparing to the alone graphene similar system.    

\section{Conclusion}

We have studied the superposition of exotic graphene atomic layer with a monolayer transition-metal dichalcogenide, where singlet and as well triplet superconducting correlations is induced by proximity-effect. As first, the relativistic dynamics of Dirac fermions in graphene is altered owing to inducing a sizeable orbital band gap and resulting spin-orbit coupling. Due to the existence of two type atom in the honeycomb-like lattice of ML-TMDC, the sublattice symmetry breaks in graphene lattice. Therefore, introducing a proper time-reversal and particle-hole symmetries, under which the electron-like in $K$ valley is transformed to corresponding hole-like in $K'$ valley becomes an important requirement of formalism. We have introduced relating time-reversal and, particularly, particle-hole symmetry operators satisfying Dirac-Bogoliubov-de Gennes equation. Secondly, the electron-hole superconducting excitation in graphene-TMDC hybrid structure has been investigated. Indeed, existence of significant intrinsic and Rashba type spin-orbit couplings (even pseudospin inversion asymmetry (PIA), which is negligible due to its very small influence) in graphene-TMDC has led to emerging distinct superconducting subgap. This effect can be revealed by considering Andreev reflection process at the normal-superconductor interface for the incident electrons with energy below the superconductor subgap energy. This point of view is proposed as a future research. We have obtained that both Dirac-point gap and effective superconductor Fermi-point subgap can be controlled by tuning the SOC features of TMDC. It has been found that, in the absence of all SOC terms, the value of superconducting effective subgap almost vanishes. The location of the superconducting effective gap is adjusted by values of superconductor chemical potential. Notice that in superconducting graphene-$WS_2$ hybrid, the absence of all SOC terms has given rise to strongly increasing Dirac-point gap. Interestingly, the Mexican hat shape of superconducting conduction band excitations in $MoS_2$ has been converted to Cowboy hat in graphene-$WSe_2$, which means that inversion band regime originated from strong SOC is appeared even in superconducting electron-hole energy spectra. Also, in superconducting graphene-$WSe_2$, when all SOC terms are zero the Dirac gap remains unchanged, and the band inversion regime disappears. A key finding of the present work is that the superconducting effective subgap has been increased when induced superconductor is $p$-wave order, in comparing to that occurred in $s$-wave symmetry.


\newpage

\textbf{Figure captions}\\
\textbf{Figure 1} Sketch of graphene-ML-MDS with proximity of a superconductor electrode.\\
\textbf{Figure 2(a), (b), (c), (d)} (Color online) The energy spectrum of graphene-TMDC as a function of wave vector in the presence of intrinsic and Rashba spin-orbit coupling. Red solid lines indicate spin up, and blue dashed lines distinguish spin down. Graphene on (a) $MoS_2$, (b) $MoSe_2$  and (c) $WS_2$ possess conventional band structure, while graphene-$WSe_2$ (d) indicates a band inversion. \\
\textbf{Figure 3(a), (b), (c), (d)} (Color online) The energy dispersion in superconductor graphene-TMDC is plotted versus  wave vector, when $\Delta_0=0.05\ meV$, $\mu_S=2\ meV$. The plots show the effect of $\lambda_R$ and $\lambda_I^{A,B}$ terms on the excitation energy. The superconducting effective subgap of graphene on (c) $WS_2$  and (d) $WSe_2$ is larger than that of graphene on (a) $MoS_2$  and (b) $MoSe_2$.\\
\textbf{Figure 4} (color online) The energy dispersion in superconductor graphene-$WSe_2$ is shown as a function of $k_x$ and $k_y$. We set the magnitude of all SOC terms from table 1, and $\Delta_0=0.2 \ meV$, $\mu_S=-2.1 \ meV$. \\
\textbf{Figure 5(a), (b), (c), (d)} (color online) The energy dispersion in superconductor graphene-TMDC are presented versus wave vector for different values of chemical potential. Dependence of excitation energy is shown for three different values $\mu_S=2\ meV$ (black solid line), $5\ meV$(red dashed line) and $10\ meV$ (blue dashed line), where $\Delta_0=0.05\ meV$. Panel (a) presents graphene-$MoS_2$, Panel (b) graphene-$MoSe_2$, Panel (c) graphene-$WS_2$ and Panel (d) graphene- $WSe_2$.\\
\textbf{Figure 6(a), (b)} (color online) The plot shows, particularly, superconducting effective subgap in the cases of spin-singlet $s$-wave and spin-triplet $p$-wave symmetries. We set the $\Delta_0=0.2\ meV$ and $\mu_S=2.1\ meV$. Dashed and solid lines distinguish $s$-wave and $p$-wave paring, respectively. Panel (a): graphene on $MoS_2$ and $MoSe_2$. Panel (b): graphene on $WS_2$ and $WSe_2$. 

\newpage

\begin{table}[h]
\begin{center}
\begin{tabular}{l |*{6}{c}r}
TMDC              & $t(eV)$ & $\Delta$ & $\lambda_I^A$ & $\lambda_I^B$ & $\lambda_R$ & $\lambda_{PIA}^A$ & $\lambda_{PIA}^B$ \\ \hline
\hline
$MoS_2$           & 2.668 & 0.52 & -0.23 & 0.28 & 0.13 & -1.22 & -2.23 \\
$MoSe_2$          & 2.526 & 0.44 & -0.19 & 0.16 & 0.26 & 2.46 & 3.52 \\
$WS_2$            & 2.657 & 1.31 & -1.02 & 1.21 & 0.36 & -0.98 & -3.81 \\
$WSe_2$           & 2.507 & 0.54 & -1.22 & 1.16 & 0.56 & -2.69 & -2.54 \\
\end{tabular}
\caption{The values of the band-structure parameters for graphene on $MoS_2$, $MoSe_2$, $WS_2$ and $WSe_2$ $\cite{MG}$. All values of parameters are in $meV$, except nearest-neighbor hopping $t$.}
\end{center}
\end{table}

\begin{figure}[p]
\epsfxsize=0.5 \textwidth
\begin{center}
\epsfbox{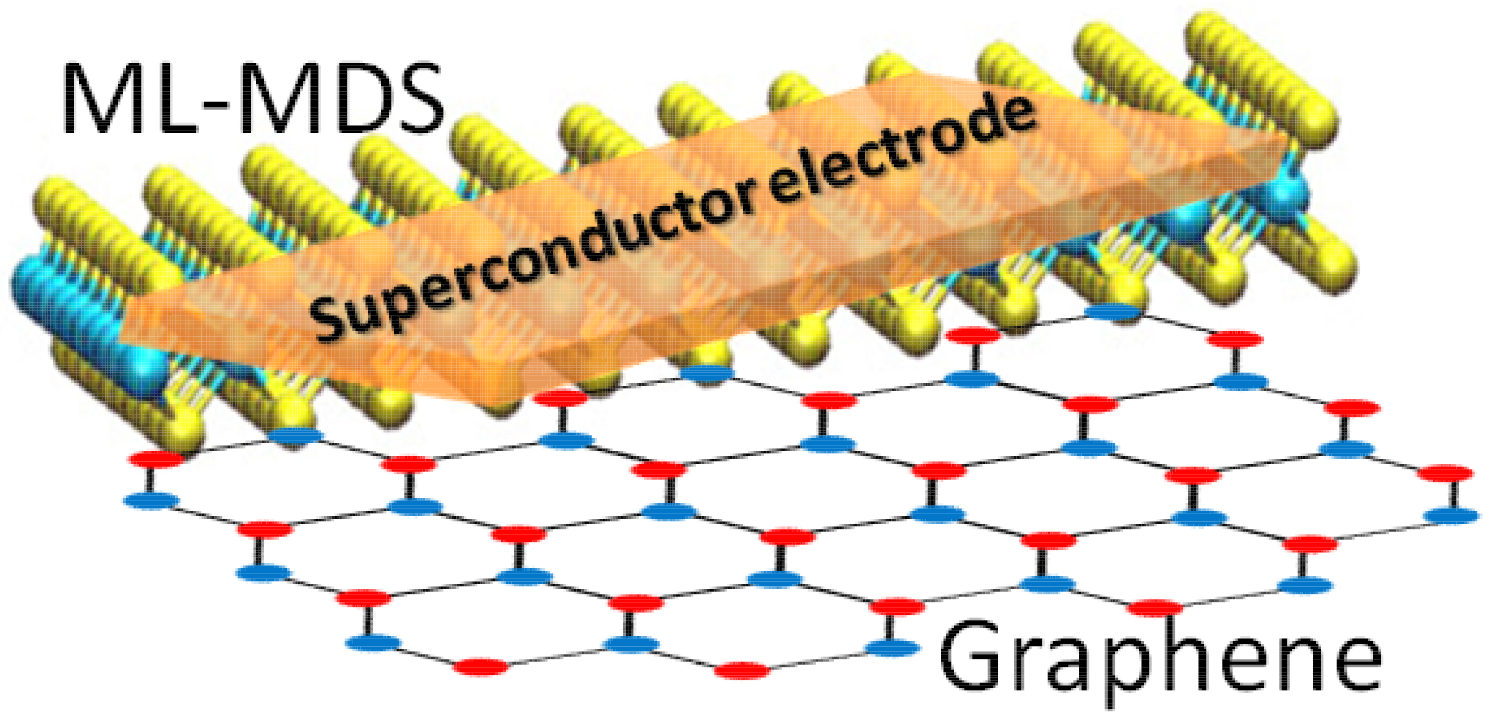}
\setcounter{figure}{0}
\caption{\footnotesize }
\end{center}
\end{figure}

\begin{figure}[p]
\epsfxsize=0.6 \textwidth
\begin{center}
\epsfbox{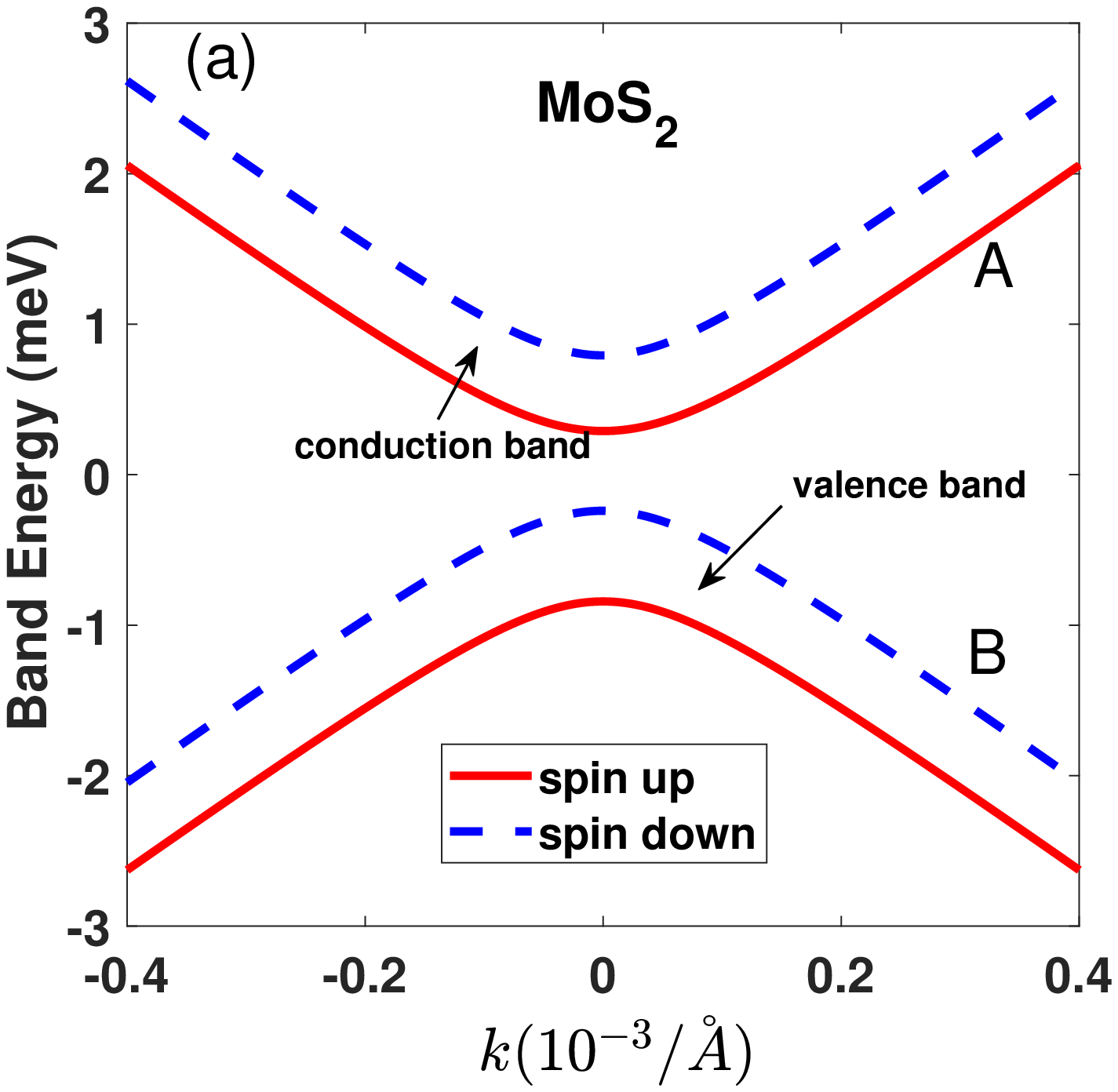}
\end{center}
\end{figure}

\begin{figure}[p]
\epsfxsize=0.6 \textwidth
\begin{center}
\epsfbox{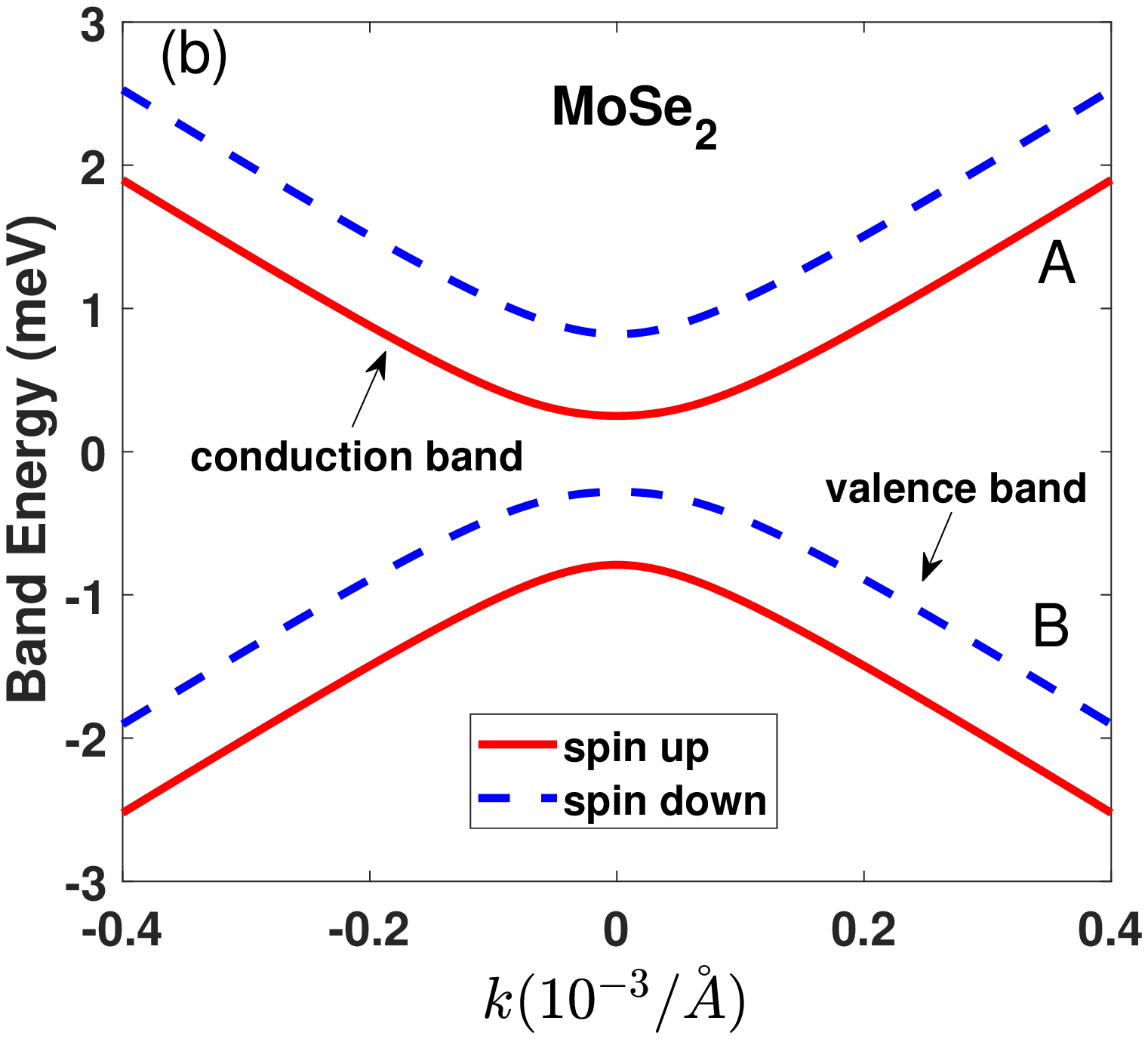}
\end{center}
\end{figure}

\begin{figure}[p]
\epsfxsize=0.6 \textwidth
\begin{center}
\epsfbox{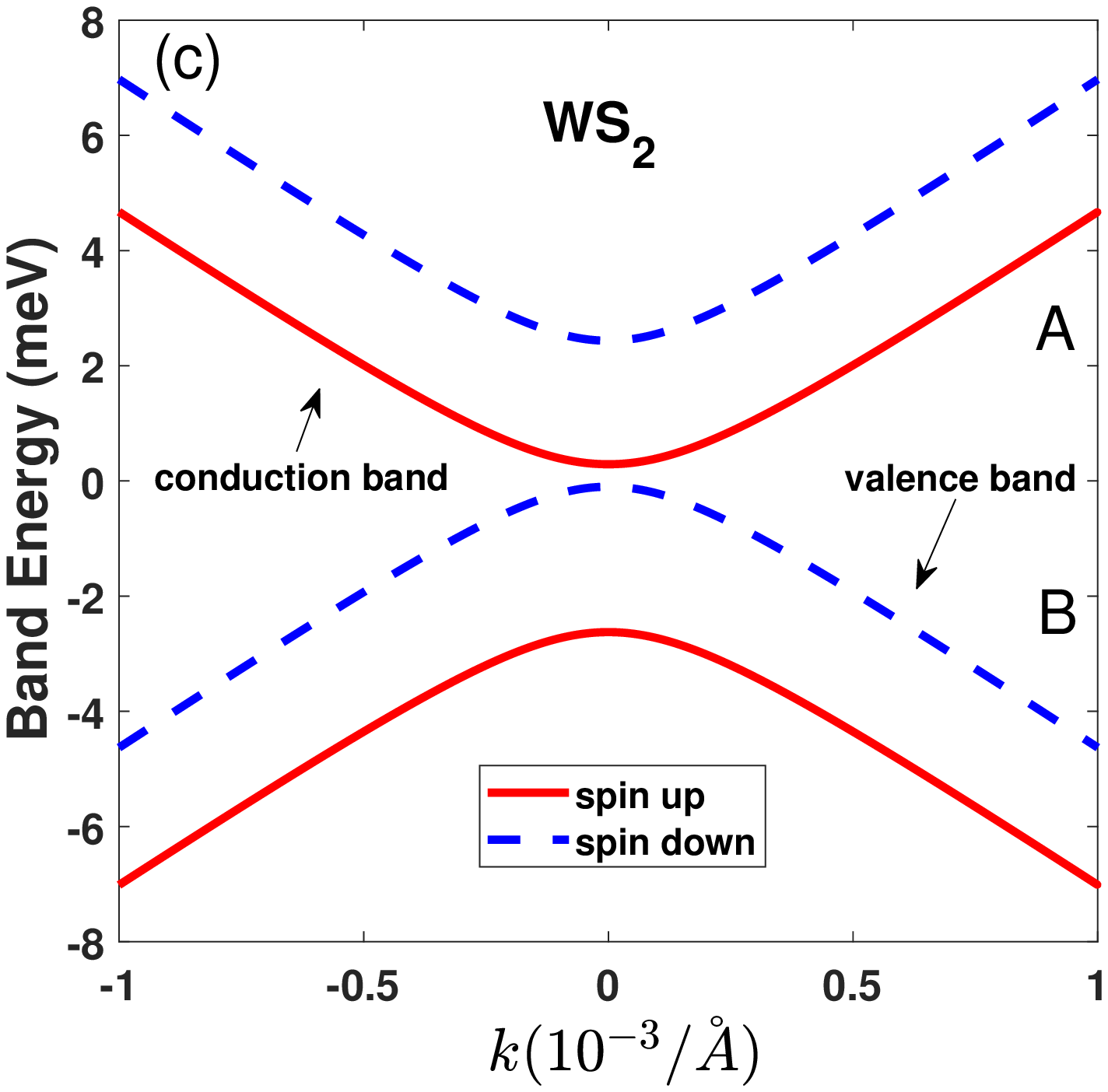}
\end{center}
\end{figure}

\begin{figure}[p]
\epsfxsize=0.6 \textwidth
\begin{center}
\epsfbox{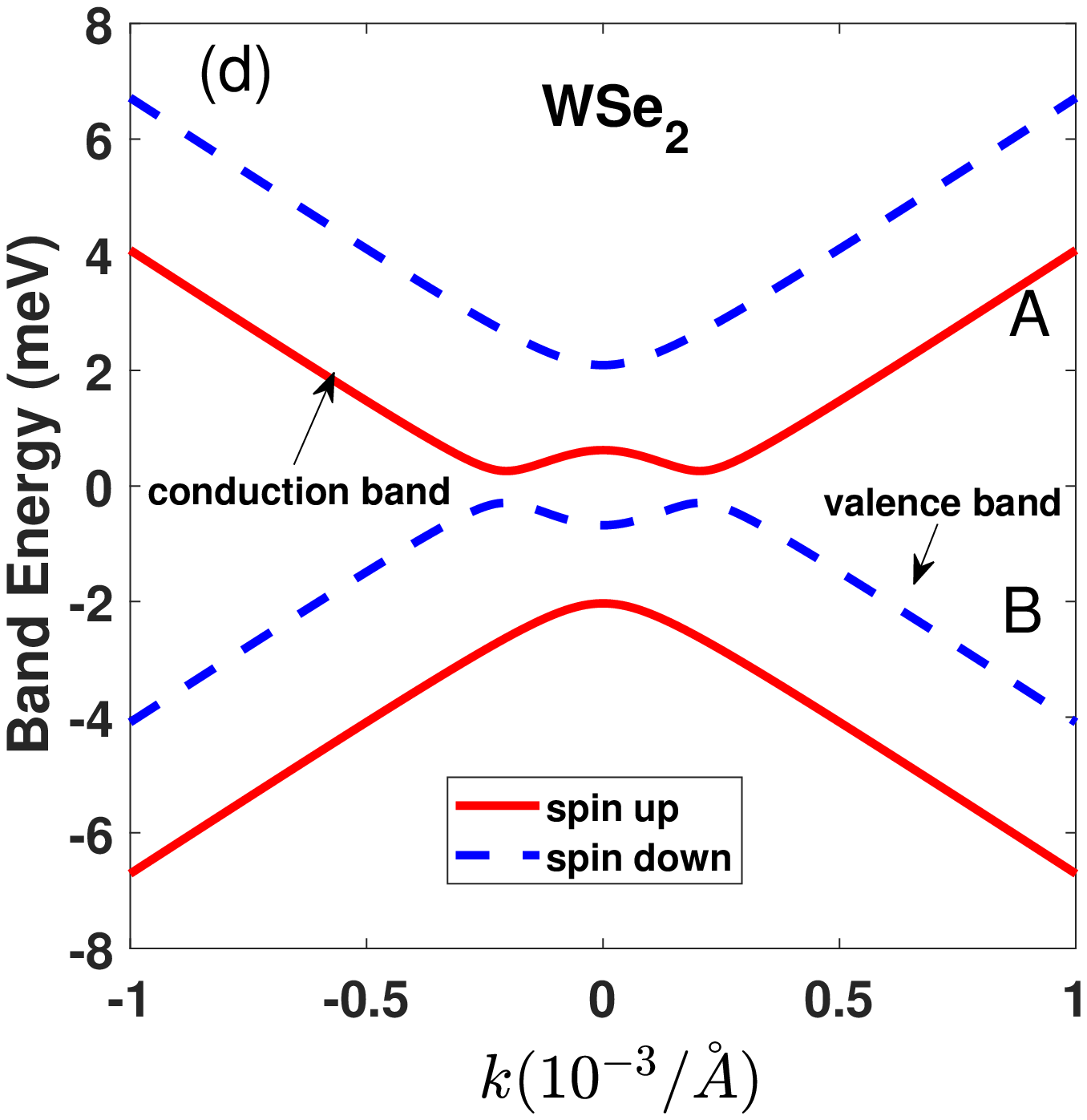}
\setcounter{figure}{1}
\caption{\footnotesize (a),(b),(c),(d)}
\end{center}
\end{figure}

\begin{figure}[p]
\epsfxsize=0.6 \textwidth
\begin{center}
\epsfbox{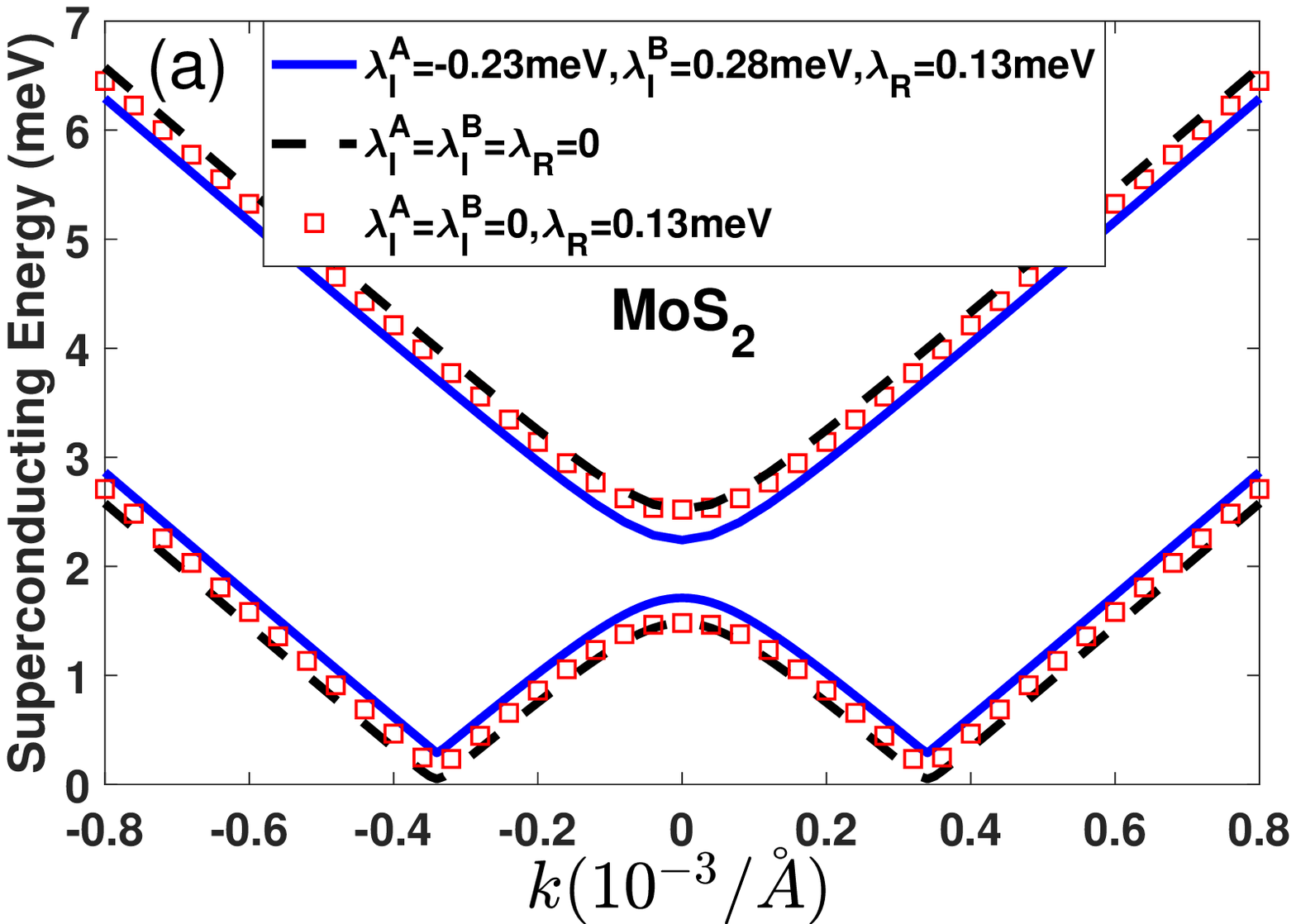}
\end{center}
\end{figure}

\begin{figure}[p]
\epsfxsize=0.6 \textwidth
\begin{center}
\epsfbox{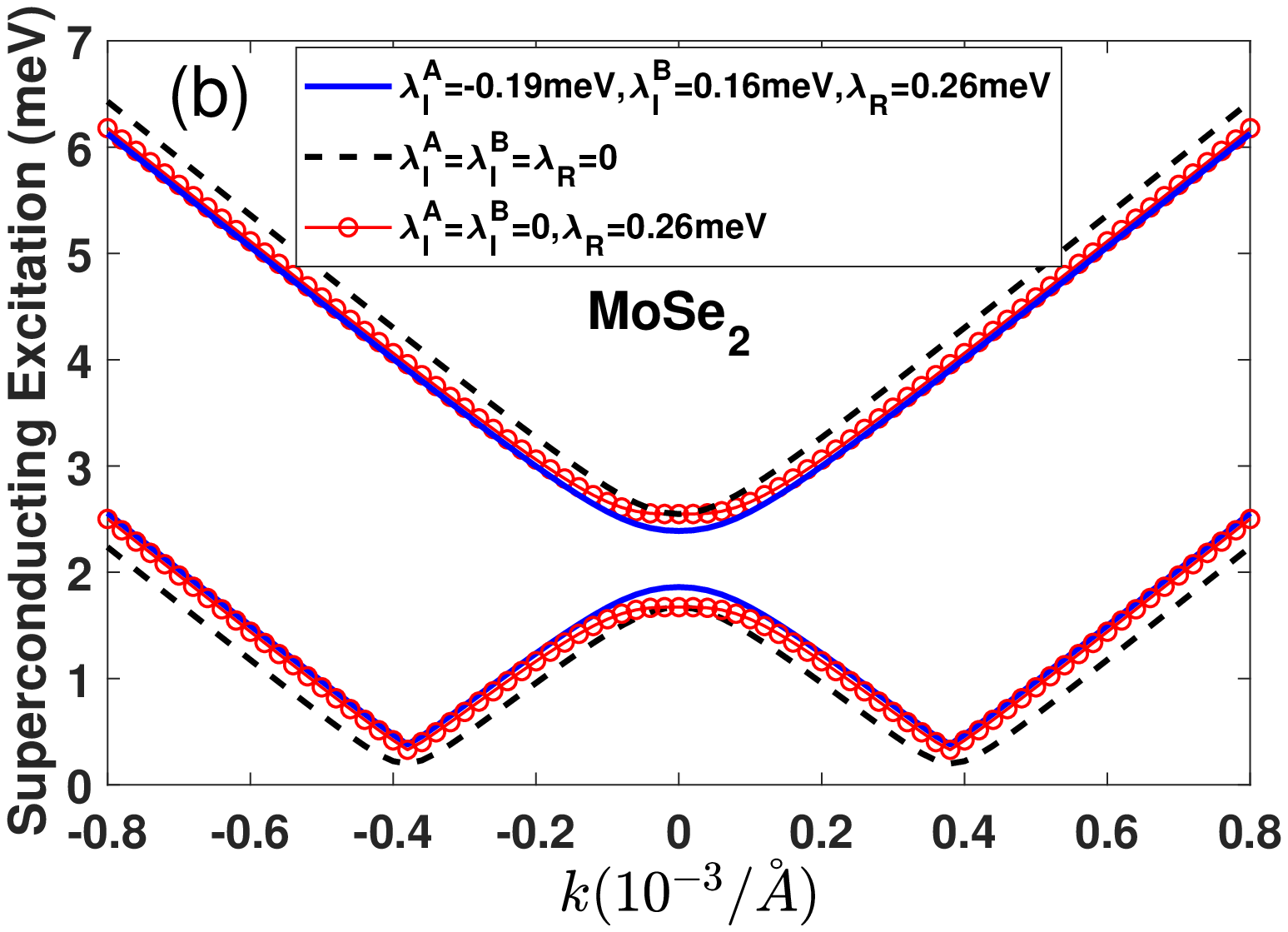}
\end{center}
\end{figure}

\begin{figure}[p]
\epsfxsize=0.6 \textwidth
\begin{center}
\epsfbox{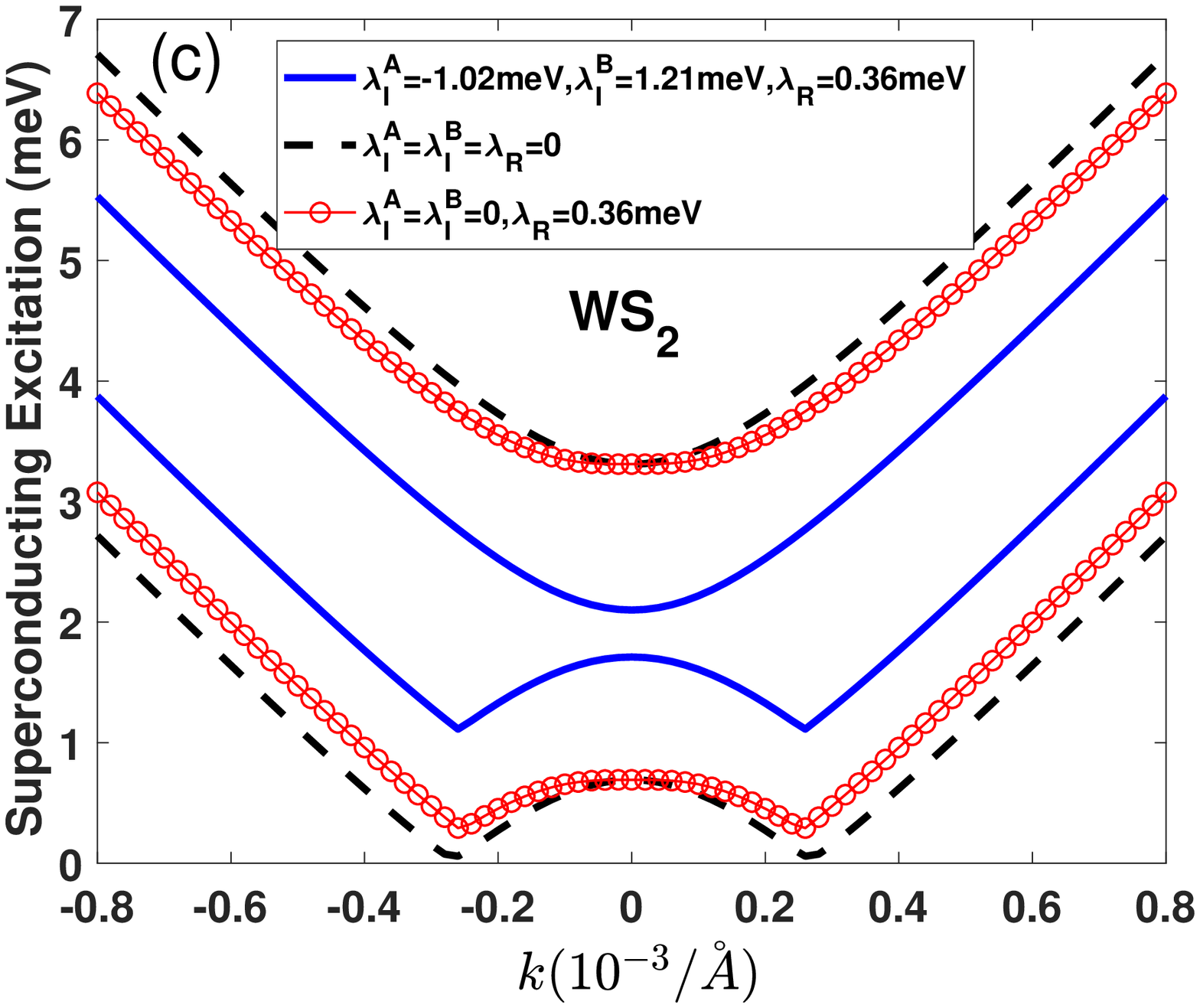}
\end{center}
\end{figure}

\begin{figure}[p]
\epsfxsize=0.6 \textwidth
\begin{center}
\epsfbox{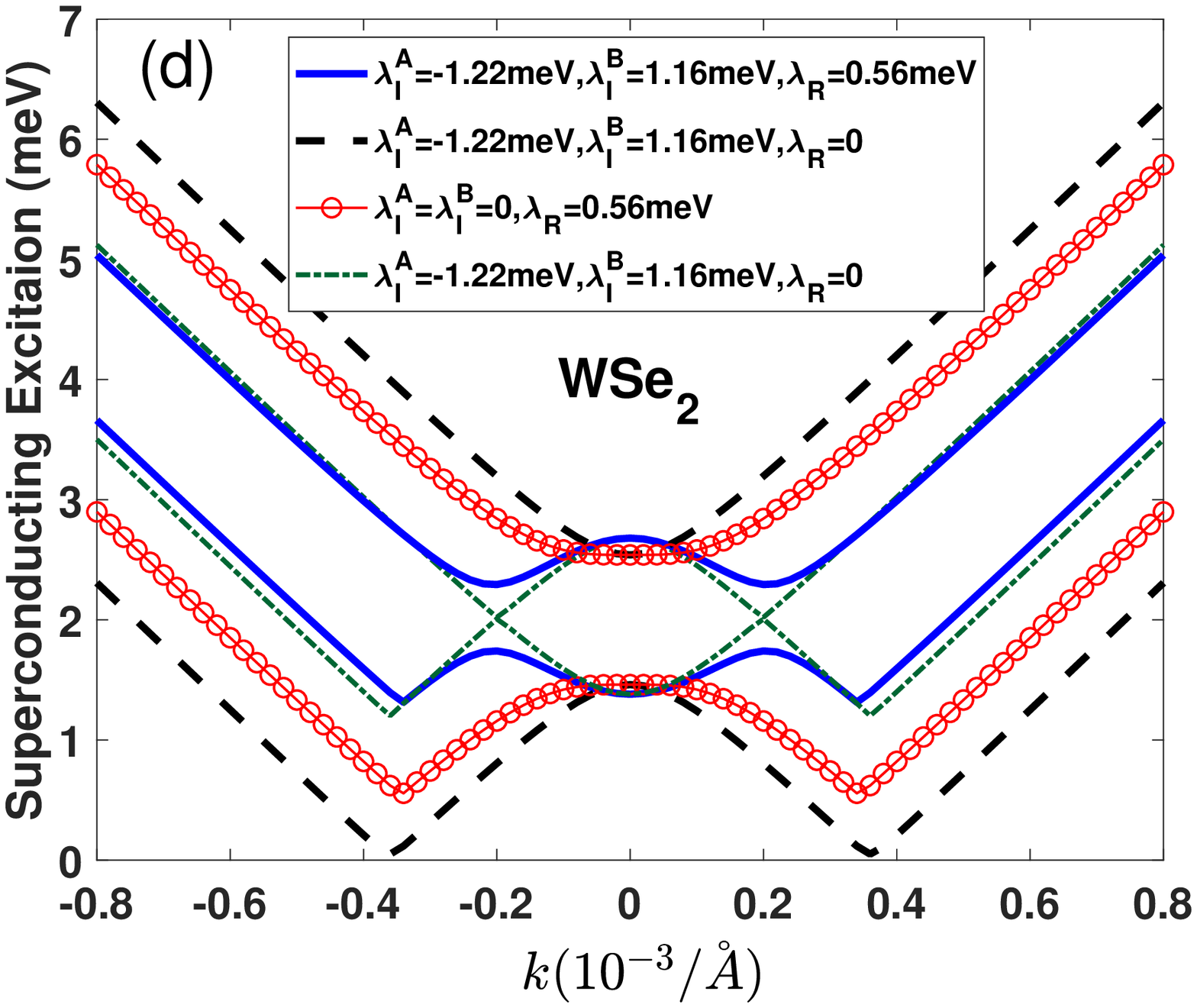}
\setcounter{figure}{2}
\caption{\footnotesize (a),(b),(c),(d)}
\end{center}
\end{figure}

\begin{figure}[p]
\epsfxsize=0.6 \textwidth
\begin{center}
\epsfbox{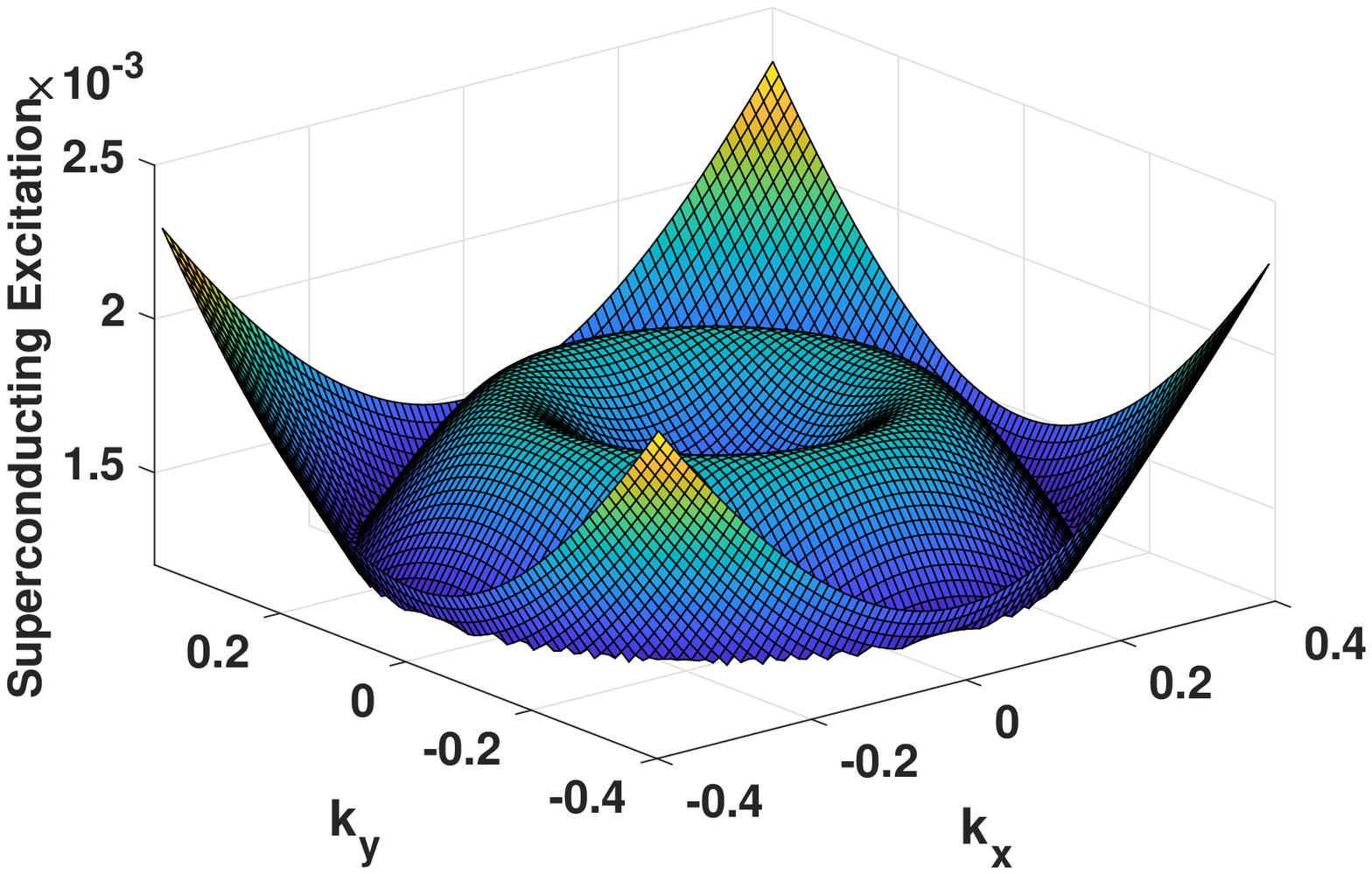}
\setcounter{figure}{3}
\caption{\footnotesize }
\end{center}
\end{figure}

\begin{figure}[p]
\epsfxsize=0.6 \textwidth
\begin{center}
\epsfbox{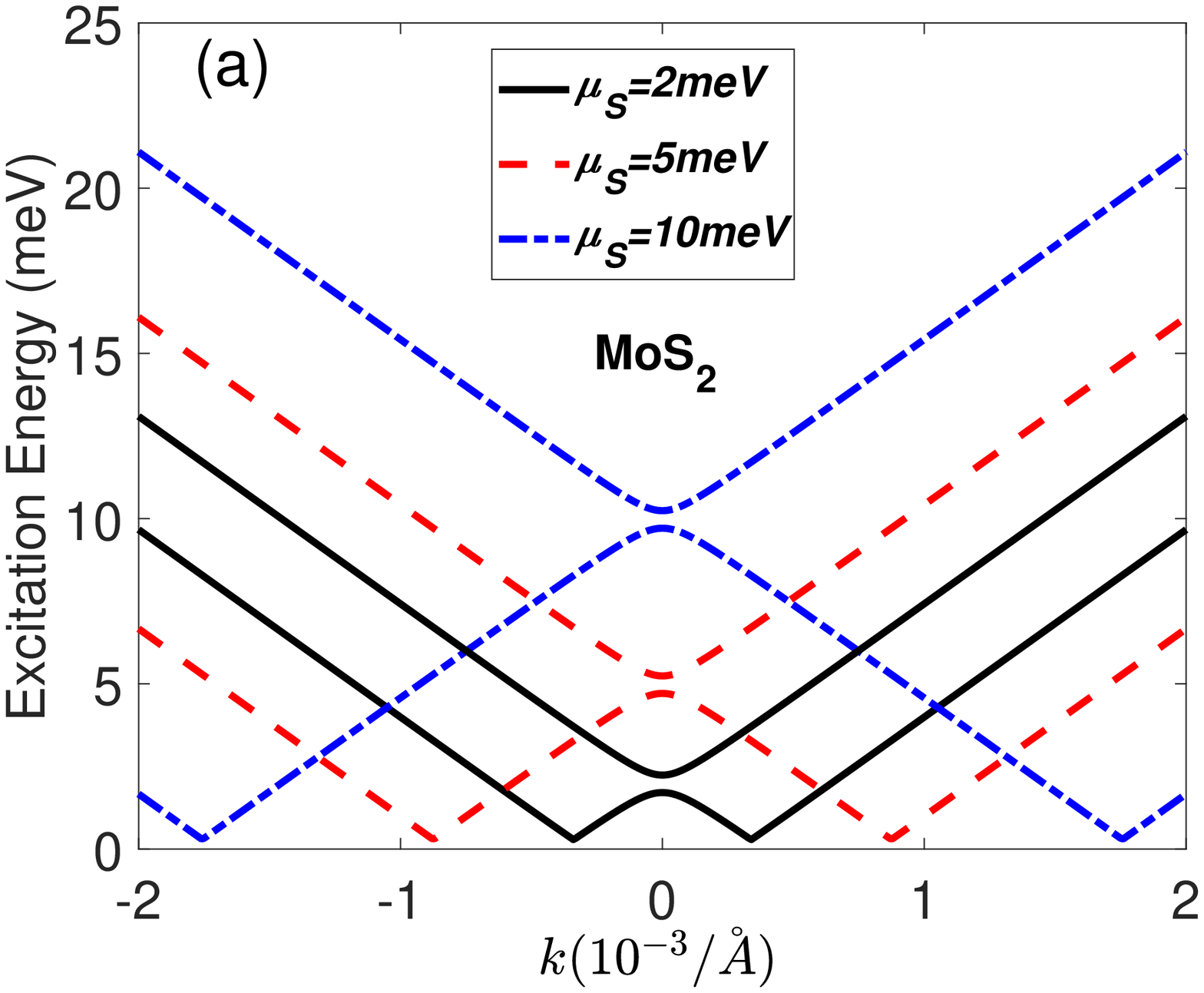}
\end{center}
\end{figure}

\begin{figure}[p]
\epsfxsize=0.6 \textwidth
\begin{center}
\epsfbox{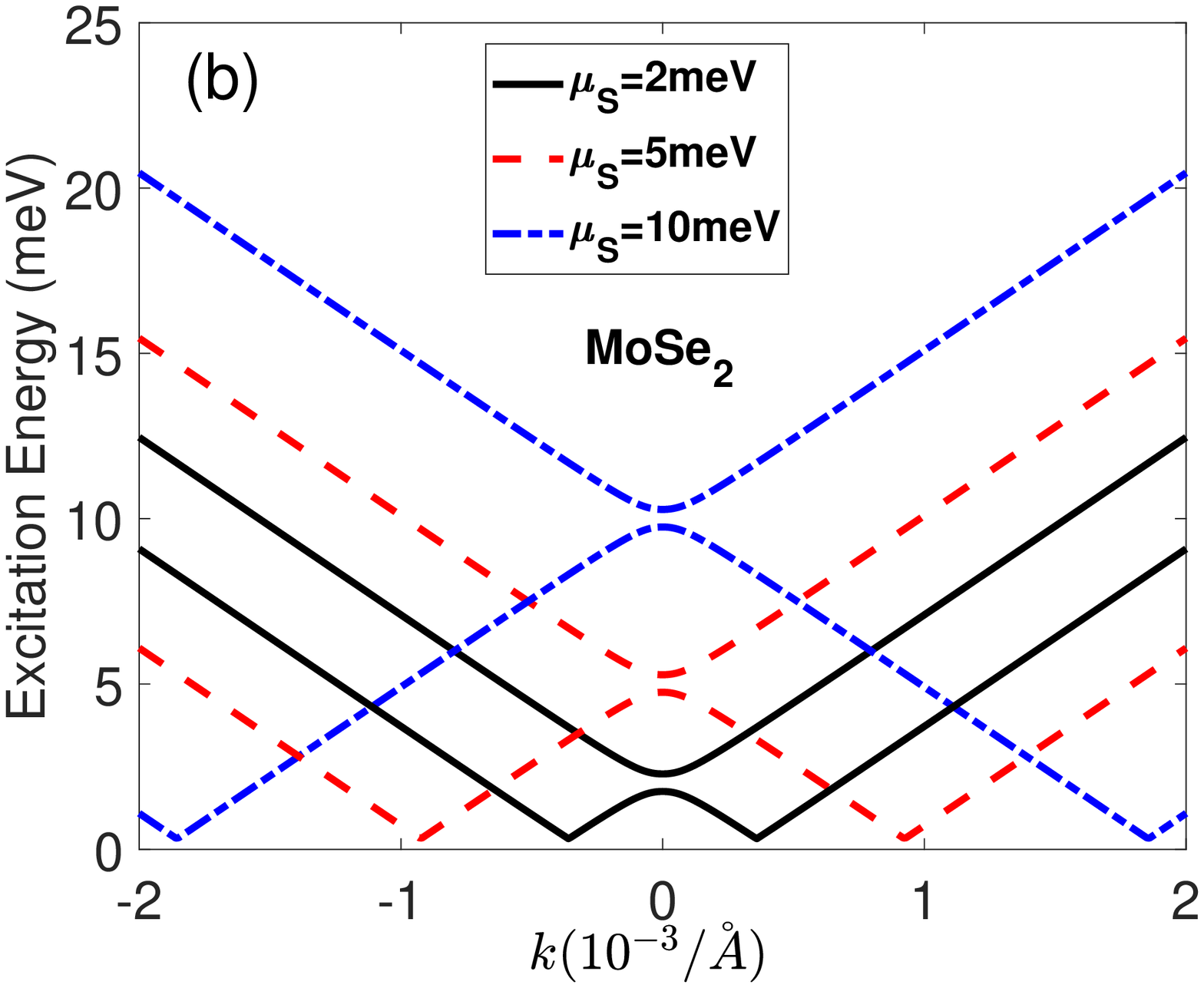}
\end{center}
\end{figure}

\begin{figure}[p]
\epsfxsize=0.6 \textwidth
\begin{center}
\epsfbox{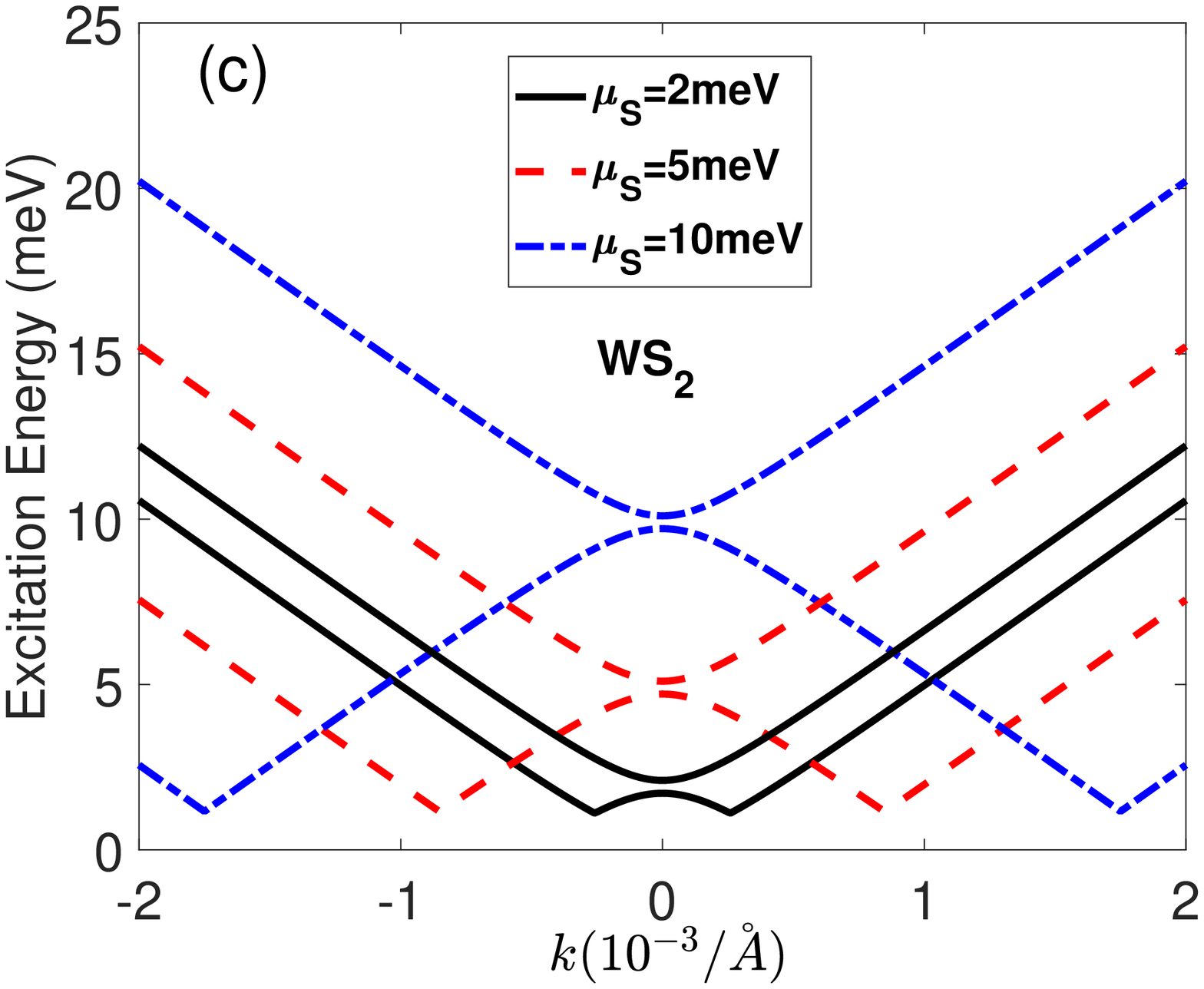}
\end{center}
\end{figure}

\begin{figure}[p]
\epsfxsize=0.6 \textwidth
\begin{center}
\epsfbox{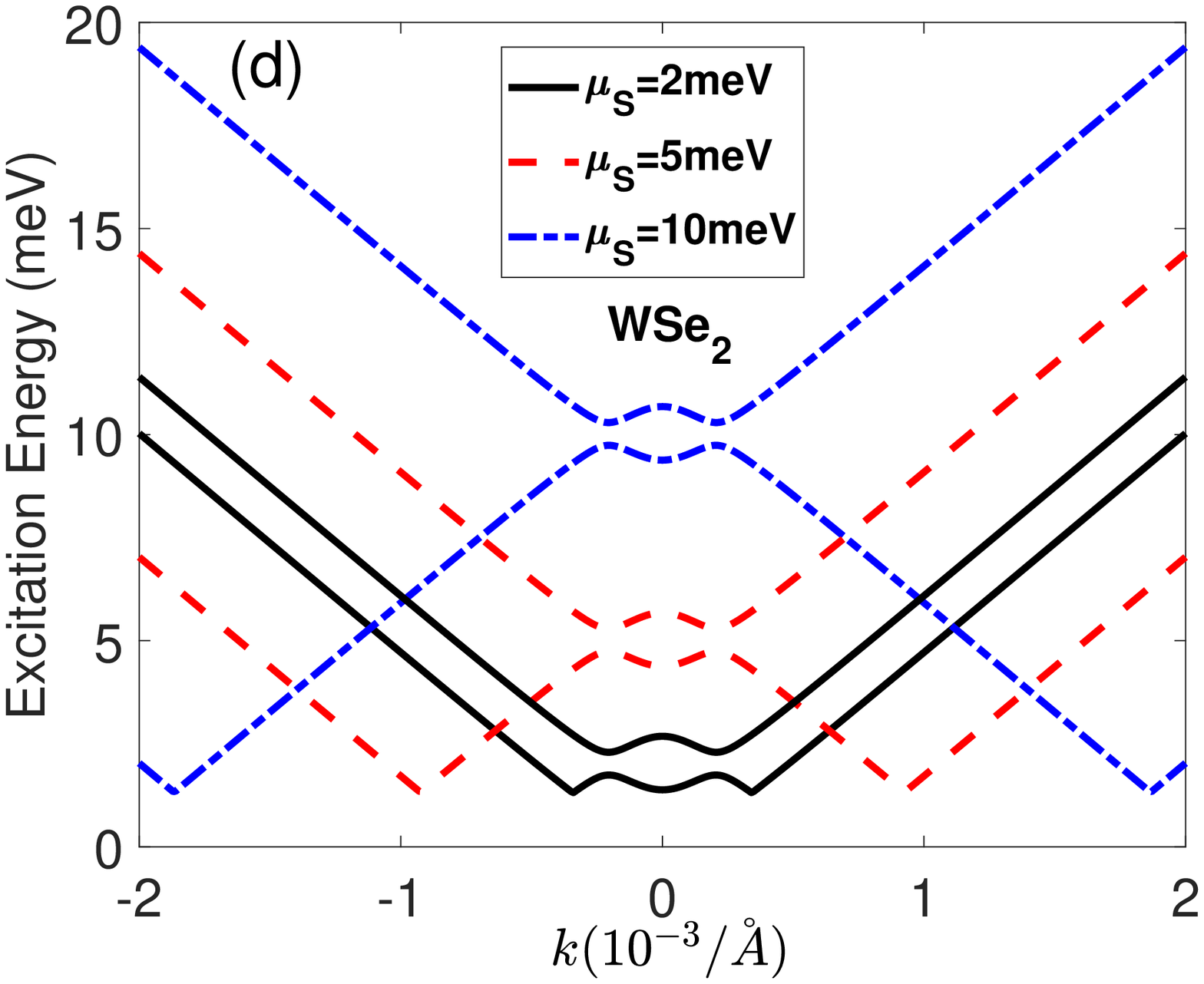}
\setcounter{figure}{4}
\caption{\footnotesize (a),(b),(c),(d)}
\end{center}
\end{figure}

\begin{figure}[p]
\epsfxsize=0.6 \textwidth
\begin{center}
\epsfbox{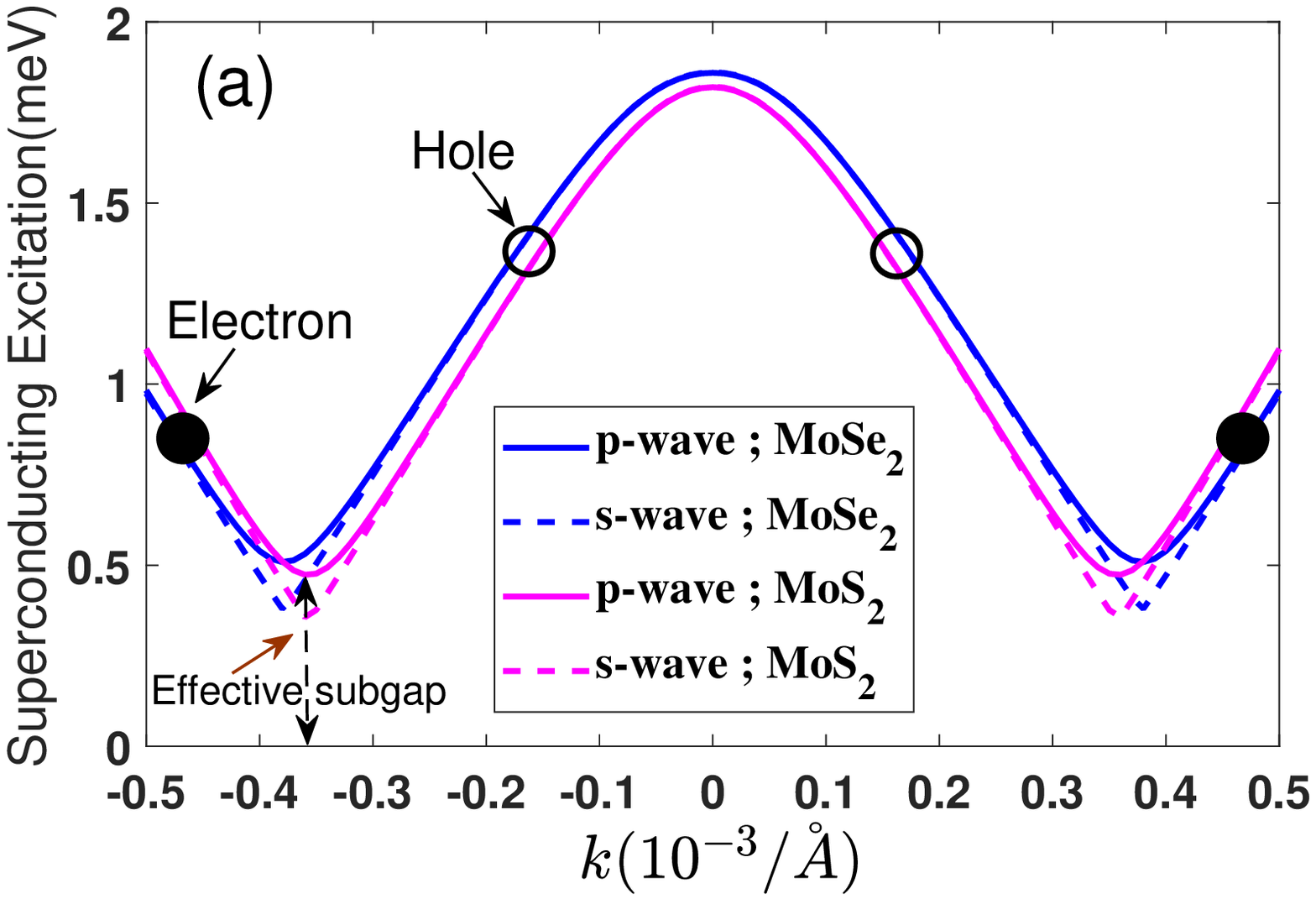}
\end{center}
\end{figure}

\begin{figure}[p]
\epsfxsize=0.6 \textwidth
\begin{center}
\epsfbox{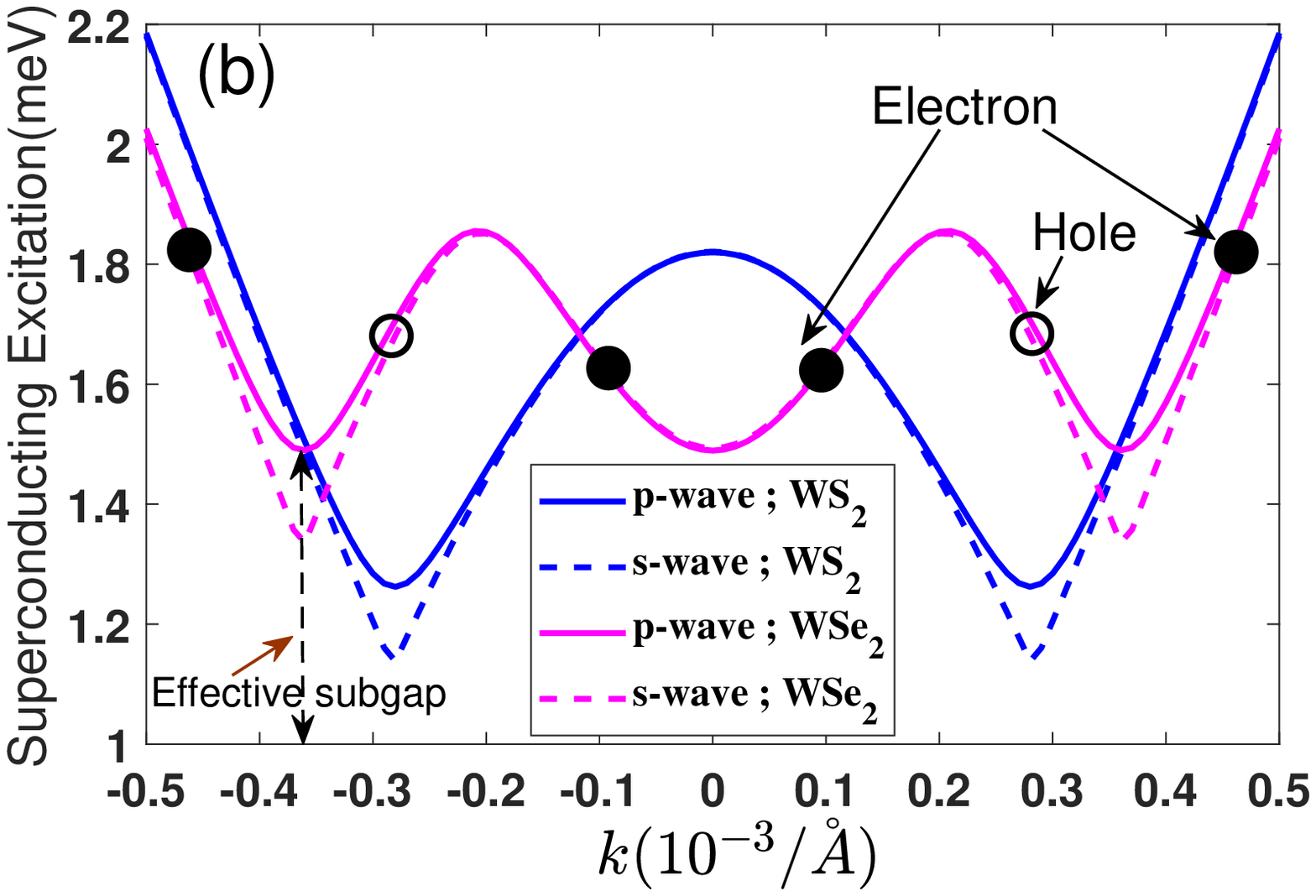}
\setcounter{figure}{5}
\caption{\footnotesize (a),(b)}
\end{center}
\end{figure}

\end{document}